\documentclass[aps, reprint,  epsfig]{revtex4-1}

\usepackage{subcaption}

\usepackage{array}
\usepackage{graphics}
\usepackage{graphicx}
\usepackage{epsfig}
\usepackage{latexsym}
\usepackage{epstopdf} 
\usepackage{color}
\usepackage{amsmath}
\usepackage{mathtools}

\usepackage{hyphenat}
\let\oldref\ref
\renewcommand{\ref}[1]{(\oldref{#1})}
\usepackage[colorlinks=true,linkcolor=blue,urlcolor=blue,citecolor=blue]{hyperref}
\usepackage{amsmath}
\usepackage[normalem]{ulem}

\def \mc {\multicolumn}

\begin{document}
\title{How attractive and repulsive interactions 
affect structure ordering and dynamics of glass-forming liquids?}

\author{Ankit Singh}
\author{Yashwant Singh}
\affiliation{Department of Physics, Banaras Hindu University, Varanasi-221 005, India.}

\date{\today}

\begin{abstract}
The theory developed in our previous papers is applied in this paper to investigate the dependence of
 slowing down of dynamics of glass-forming liquids on the 
attractive and repulsive parts of intermolecular interactions. 
Through an extensive comparison of the behavior of a Lennard-Jones
glass-forming liquid and that of its WCA reduction to a model with truncated pair potential
without attractive tail, we demonstrate why the two systems exhibit very different dynamics
despite having nearly identical pair correlation functions. In particular, we show that 
local structures characterized by number of mobile and immobile particles around a central
particle markedly differ in the two systems at densities and temperatures where their dynamics show large difference and nearly identical where dynamics nearly overlap. We also show how the parameter $\psi(T)$ that measures the role of fluctuations
embedded in the system on size of the cooperatively reorganizing cluster (CRC)
and the crossover temperature $T_{a}$ depend on the intermolecular interactions. These parameters stemming from the intermolecular interactions
characterize the temperature and density dependence of structural relaxation time
$\tau_{\alpha}$. The quantitative and qualitative agreements found with simulation results for the two systems
suggest that our theory brings out the underlying features that determine
dynamics of glass-forming liquids.

\end{abstract}

\pacs{64.70.Q-, 61.20.Gy, 64.70.kj}

\maketitle

\section{Introduction\label{Intro}}
In previous papers \cite{Ankit,ASingh, Singh} of this series we introduced a new theory for slowing 
down of dynamics of fragile liquids on cooling. The theory identifies the local structural order
that defines the cooperativity of relaxation and brings forth a temperature $T_{a}$ and a 
temperature dependent parameter $\psi(T)$ that characterize temperature and density dependence of structural
relaxation time. In Ref.~\cite{Singh}, hereafter referred to as $I$, we reported results for the
Kob-Anderson $80:20$ binary Lennard-Jones (LJ) mixture \cite{LJ} and compared them with simulation results \cite{Tarjus_PRL,Tarjus_JCP}.
In this paper we investigate how different branches of intermolecular interaction potentials affect
the local structural order and 
dynamics of supercooled liquids. In particular, we consider a model system in which particles
interact via a purely repulsive pair potential formed by truncating the LJ potential at its minimum \cite{Weeks}; a potential
known as the Weeks-Chandler-Anderson (WCA) binary mixture potential.
The two potentials, the LJ
and WCA, have same repulsive core but different attractive part; the LJ has an attractive tail while the WCA has no attraction. These models therefore offer an ideal benchmark for evaluating 
the role of repulsive and attractive interactions on the local structural order and on slowing down of 
dynamics of glass-forming liquids, and have been studied extensively over last several years
\cite{Tarjus_PRL,Tarjus_JCP,LBEuro,LGMC,Toxvaerd,Pedersen,DC_PRE,DC,Hocky,Schweizer,Banerjee,MKNandi,ABanerjee,Chattoraj,Tong,Landes}.
We compare results of the two systems and identify causes that give rise a large difference in
their local structural order and dynamics at lower liquid-like densities but almost identical
results at sufficiently large densities.  

It is almost universally accepted that the structure and thermodynamics of nonassociated liquids 
are primarily governed by
the short range repulsive branch of pair potential while the attractive part of the potential
provides, in the first approximation, a homogeneous cohesive background \cite{Weeks,Chandler,Hansen}. This led to
formulation of theories in which properties of liquids are related to those of the repulsive
core reference system, the attractive part of the potential being treated as a perturbation. The success of these perturbation theories \cite{Hansen} in predicting equilibrium properties of normal
liquids led to expectation that the structure and dynamics of supercooled liquids should also
be governed primarily by the local packing and the steric effects produced by the repulsive 
forces \cite{Bembenek,Young}. This expectation was, however challenged by results found via molecular dynamics
simulations by Bertheir and Tarjus \cite{Tarjus_PRL,Tarjus_JCP} for the LJ and WCA mixtures. They found that while 
these systems exhibit nearly identical equilibrium structure, but at lower liquid-like densities and temperatures,
very different dynamics: The value of structural relaxation time $\tau_{\alpha}$ is much smaller 
in the WCA system than in the LJ ones at the same supercooled temperature indicating that the attractive
forces have a nonperturbative effect on the relaxation dynamics. However, this large difference
in values of $\tau_{\alpha}$ decreases and ultimately vanishes as the density is significantly 
increased. Recent simulation studies \cite{Chattoraj,Tong} done in the two  and in three dimensions show similar
effect of attractive interactions in slowing down of dynamics in the supercooled region.

The above findings bring forth the limitations \cite{LBEuro,LGMC} of ``microscopic'' theories based on pair correlation
function $g(r)$ \cite{Gotze,Kirkpatrick,Schweizer_JCP}. For example, mode coupling theory was found to largely underestimate the 
difference in dynamics of the WCA and LJ systems \cite{LGMC}. This negative result led some authors to believe
that $g(r)$ does not contain the physical information required to capture subtleties involved
in dynamical slowdown and therefore any theory based on $g(r)$ is bound to fail \cite{LBEuro}.
It was suggested that the origin of this failure might be their neglect of higher order than
pair correlations. Subsequent simulations of triplet correlation functions indicate 
that the local ordering is more pronounced in the LJ system, an observation consistent with its
slower dynamics \cite{DC_PRE}. This pronounced structure was, however, shown to arise due, to a good
extent, to an amplification of the small discrepancies observed at the pair level \cite{DC}.
Recently, Landes \textit{et al.} \cite{Landes} have used a standard machine learning algorithm to show that
a properly weighted integral of $g(r)$ which amplifies the subtle differences between the 
two systems, correctly captures their dynamical differences. Local structure analysis using
topological cluster classification \cite{Taffs} and Voronoi face analysis \cite{DC_PRE}
also point to subtle structural difference. Using the configurational entropy as the thermodynamic maker via the Adam-Gibbs relation \cite{Adam}, Banerjee \textit{et al.} \cite{Banerjee,MKNandi,ABanerjee}
showed that the difference in 
dynamics of the WCA and LJ systems is due to difference in their configurational entropy.
These results prompt one to ask whether a unified physical framework exists to understand the
structure and dynamics and their relationship for systems such as the LJ and WCA glass forming liquids. Our goal here is to find answer to this question.

In Sec.~\ref{Theory} we give a brief account of our theory relevant to the present work.
In Sec.~\ref{3} we calculate and report results for the WCA system and compare them with those
found for the LJ system reported in $I$. Sec.~\ref{Conclusion} is devoted to conclusion 
that emerged from
similarity and contrast of results of the two systems
and comparison with simulation findings.

\section{Theory: \label{Theory}}
 
\begin{figure*}[t]
\includegraphics[scale=0.6]{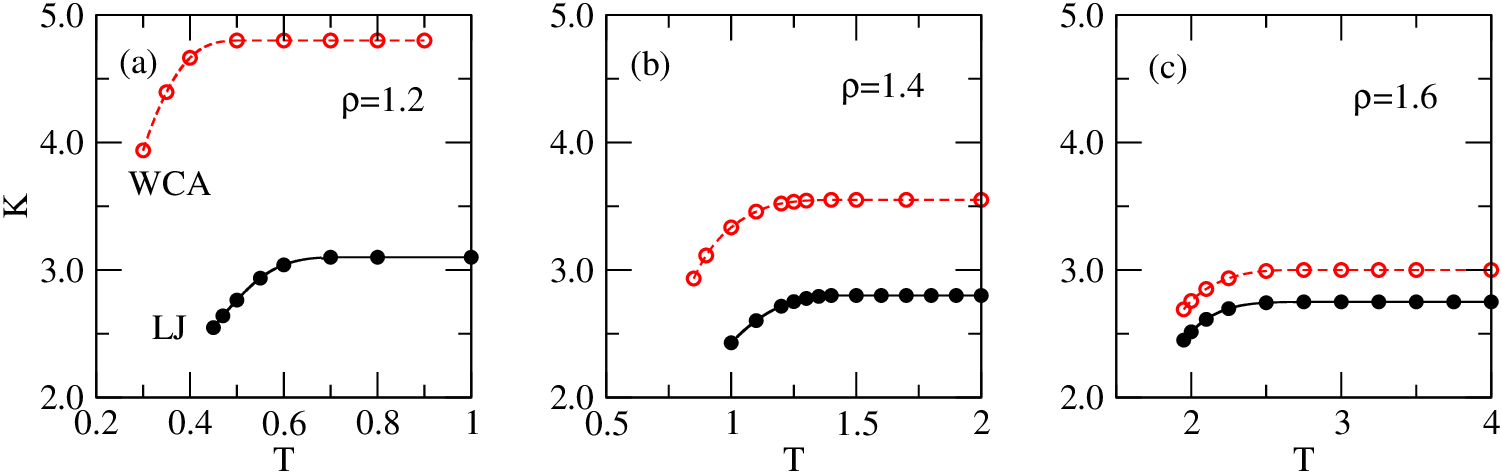}
\caption{Comparison of values of $K$ for the LJ and WCA systems.
These values were found when value of $\psi$ was fixed to $1$
at all temperatures. In both cases, $K$ is constant above a density dependent temperature, $T_{a}$, and decreases from its constant value on cooling below $T_{a}$. Values of $T_{a}$ of the two systems
at different densities are given in Table~\ref{LJ-Data}. Symbols represent calculated values and curves are least square fit.}
\label{fig-1}
\end{figure*}

Our theory provides a method to distinguish and calculate number of dynamically free,
metastable and stable neighbors of a tagged (central) particle in a liquid at different
temperatures and densities. This is achieved by including momentum distribution in the 
definition of $g(r)$ and using information of the configurational entropy $S_{c}$.
The $g(r)$ of a binary mixture written in the center-of-mass coordinates is \cite{Ankit,ASingh, Singh}, 
\begin{equation}
 g_{\alpha\gamma}(r)=\left(\frac{\beta}{2\pi\mu}\right)^{\frac{3}{2}}\int \mathrm{d}{\bf p} \  
\mathrm{e}^{-\beta (\frac{p^2}{2\mu} + w_{\alpha\gamma}(r))} . 
\end{equation}
Here $\beta=(k_{B}T)^{-1}$ is the inverse temperature measured in units of 
the Boltzmann constant $k_{B}$, ${\bf p}$  is the relative momentum of a particle of mass 
$\mu=m/2$ ($m$ being the mass of a particle of the liquid). The potential of mean force $w_{\alpha\gamma}(r)=-k_{B}T\ln g_{\alpha\gamma}(r)$ is a sum of the (bare) potential energy $u_{\alpha\gamma}$  and the system induced potential energy of interaction between a pair of particles 
of species $\alpha$ and $\gamma$ separated by distance $r$ \cite{Hansen}.
The peaks and troughs of $g_{\alpha\gamma}(r)$ create, respectively, minima and maxima in $\beta w_{\alpha\gamma}(r)$
as shown in Fig.~\ref{fig-1} of $I$.
A region between two maxima, leveled as $i-1$ and $i$ $(i\geq 1)$ is denoted as $i$th shell
and minimum of the shell as $\beta w_{\alpha\gamma}^{(id)}$.
The value of $i$th maximum (barrier) is denoted as $\beta w_{\alpha\gamma}^{(iu)}$ and its location by $r_{ih}$.

All those particles in region of $i$th shell whose energies are 
less or equal to the barrier, $w_{\alpha\gamma}^{(iu)}$ would be trapped as they do not have 
enough energy to escape the barrier.
These particles are considered as bonded (nonchemical) with the central particle.
On the other hand, all those particles whose energies are higher than 
$\beta w_{\alpha\gamma}^{(iu)}$ are free to move around and collide with other 
particles.
The number of bonded particles is found form a part of
$g_{\alpha\gamma}(r)$ defined as \cite{Ankit, Singh}

\begin{eqnarray}\nonumber
g_{\alpha\gamma}^{(ib)}(r)  &=& 4\pi(\frac{\beta}{2\pi\mu})^{3/2} \mathrm{e}^{-\beta w_{\alpha\gamma}^{(i)}(r)} 
\int_{0}^{\sqrt{2\mu[w_{\alpha\gamma}^{(iu)}-w_{\alpha\gamma}^{(i)}(r)]}}      \\
&&  \times \mathrm{e}^{-\beta p^2/2\mu} p^2 \mathrm{d}p ,
\end{eqnarray}
where $w_{\alpha\gamma}^{(i)}(r)$ is the effective potential in the range of $r_{il}\leq r \leq r_{ih}$
of $i$th shell. Here $r_{il}$ is value of $r$ where $w_{\alpha\gamma}^{(i)}(r)= w_{\alpha\gamma}^{(iu)}$
on the left hand side of the shell (see Fig.~\ref{fig-1} of $I$). The total number of particles that 
form  bonds with the central particle of species $\alpha$ is
\begin{equation}
n_{\alpha}^{(b)} = 4\pi \sum_{i} \sum_{\gamma}\rho_{\gamma}\int_{r_{il}}^{r_{ih}} 
g_{\alpha\gamma}^{(ib)}(r) r^2 \mathrm{d}r  ,
\end{equation}
where summations are over all shells and over all species and $\rho_{\gamma}$ is number density
of $\gamma$ component. The part of $g_{\alpha \gamma}(r)$ that corresponds to free particles is 
$g_{\alpha \gamma}^{(f)}(r)=g_{\alpha \gamma}(r)-g_{\alpha \gamma}^{(b)}(r)$.

Fluctuations embedded in the system (bath) activate some of these bonded particles, particularly those
whose energies are close to the barrier height, to escape the shell.  These particles are referred to as
metastable (or $m$-) particles. The remaining bonded particles stay trapped in shells
and form stable (long lived) bonds with the central particle.
They are referred to as $s$-particles.
This separation between the $m$- and $s$-particles is achieved via a parameter $\psi(T)$
which measures effect of the bath in activating bonded particles to escape the potential
barrier.
All those particles of $i$th shell whose energies lie between 
$\beta w_{\alpha\gamma}^{(i u)}-\psi$ and  $\beta w_{\alpha\gamma}^{(i u)}$ escape the shell
are $m$-particles. On the other hand, all those particles whose energies are lower than 
$[\beta w_{\alpha\gamma}^{(i u)}-\psi]$  remain trapped in the shell, are $s$-particles.
The value of $\psi(T)$ in a normal (high temperature) liquid is one but decreases on cooling below
a certain temperature denoted as $T_{a}$ which depends on density $\rho$ and on microscopic
interactions between particles.

The number of $s$-particles at a given temperature and density is found from a part of 
$g_{\alpha\gamma}(r)$ defined as
\begin{eqnarray}\nonumber
g_{\alpha\gamma}^{(is)}(r)  &=& 4\pi(\frac{\beta}{2\pi\mu})^{3/2} \mathrm{e}^{-\beta w_{\alpha\gamma}^{(i)}(r)} 
\int_{0}^{\sqrt{2\mu[w_{\alpha\gamma}^{(iu)}-\psi k_{B} T-w_{\alpha\gamma}^{(i)}(r)]}}      \\
&&  \ \times \mathrm{e}^{-\beta p^2/2\mu} p^2 \mathrm{d}p ,
\label{grs}
\end{eqnarray}
where $w_{\alpha\gamma}^{(i)}(r)$ is in the range $r''_{il}\leq r \leq r''_{ih}$. 
Here $r''_{il}$ and $r''_{ih}$ are, respectively, value of $r$ on the left and the 
right hand side of the shell where $\beta w_{\alpha\gamma}^{(i)}(r)= \beta w_{\alpha\gamma}^{(iu)}-\psi$. 
The number of $s$-particles around a $\alpha$ particle is 
\begin{equation}
n_{\alpha}^{(s)} = 4\pi \sum_{i} \sum_{\gamma}\rho_{\gamma}\int_{r''_{il}}^{r''_{ih}} 
g_{\alpha\gamma}^{(is)}(r) r^2 \mathrm{d}r  .
\label{ns}
\end{equation}

The averaged number of $s$-particle bonded with a central particle in a binary mixture is

\begin{equation}
n^{(s)} = x_{a} n_{a}^{(s)} + x_{b} n_{b}^{(s)}  ,
\label{nst}
\end{equation}
where $x_{\alpha}$ is the concentration of species $\alpha$.

The $n^{(s)}$, $s$-particles bonded with a central particle form a cooperatively reorganizing cluster 
(CRC). The number of particles in the cluster is related to the configurational entropy $S_{c}$
through the Adam and Gibbs \cite{Adam} relation,
\begin{equation}
n^{(s)}(T)+1 = \dfrac{K}{S_{c}(T)} \ \ ,
\label{Sc}
\end{equation}
where $K$ is a temperature independent constant.  
For an event of structural relaxation to take place the CRC has to reorganize irreversibly;
The energy involved in this process is the effective activation energy $\beta E^{(s)}$ of
relaxation. It is equal to the energy with which the central particle is bonded with $n^{(s)}$,
$s$-particles and is given as 
\begin{eqnarray}\nonumber
\beta E^{(s)}(T,\rho) &=& 4\pi \sum_{i}\sum_{\gamma}x_{\gamma}\rho_{\gamma} \int_{r''_{il}}^{r''_{ih}} 
[\beta w_{\alpha\gamma}^{(iu)}-\psi(T)- \beta w_{\alpha\gamma}^{(i)}(r)] \\
&& \ \times g_{\alpha\gamma}^{(is)}(r) r^2 \mathrm{d}r ,
\label{energy}
\end{eqnarray}
where energy is measured from the effective barrier 
$\beta w_{\alpha\gamma}^{(iu)}-\psi(T)$. 
The structural relaxation time $\tau_{\alpha}$ is obtained from the 
Arrhenius law,
\begin{equation}
\tau_{\alpha}(T,\rho) = \tau_{0} \exp{[\beta E^{(s)} (T,\rho)]}  ,
\label{tau_E}
\end{equation}
where $\tau_{0}$ is a microscopic time scale.

The data of $g_{\alpha \gamma}(r)$ and $S_{c}$ found from
simulations and reported in Ref.~\cite{ABanerjee} are used in the calculation.

\section{Results and Discussions \label{3}}

\begin{table}[b]
\caption{Value of constant $K$ and the crossover temperature $T_{a}$ of the LJ and WCA systems at different densities.}
\vspace{0.3cm}
\label{LJ-Data}
\begin{ruledtabular}
\begin{tabular}{c c c c c c c}

 {   } & & \mc{2}{c}{ LJ } & & \mc{2}{c}{ WCA }  \\ \cline{2-4} \cline{5-7}
$ \rho $ & & $ K $ &   $ T_{a} $  &  & $ K $ &   $ T_{a} $   \\
\hline

 $  1.2 $  &  &  $ 3.10 $  &  $ 0.68 $  & & $ 4.80 $  &  $ 0.47 $    \\
 $  1.4 $  &  &  $ 2.80 $  &  $ 1.43 $  & &  $ 3.55 $  &  $ 1.30 $   \\
 $  1.6 $  &  &  $ 2.75 $  &  $ 2.68 $  & &  $ 3.00 $  &  $ 2.68 $   \\
\end{tabular}
\end{ruledtabular}
\end{table}

\begin{figure*}[t]
\includegraphics[scale=0.6]{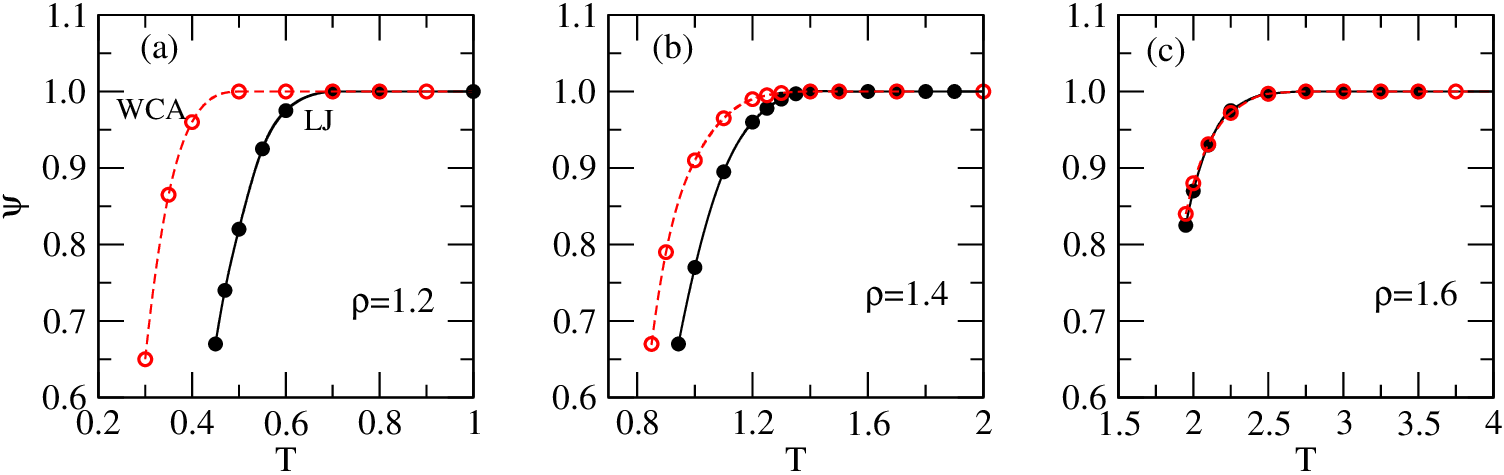}
\caption{Comparison of values of $\psi(T,\rho)$ of the two systems
at different temperatures and densities. Symbols are calculated values and curves are least square fit. A large difference in values of $\psi(T,\rho)$ of the two systems for $\rho=1.2$ below $T=0.68$ is seen whereas for $\rho=1.6$ values of $\psi(T)$ almost overlap at all temperatures.}
\label{fig-2}
\end{figure*}

\begin{figure*}
\begin{subfigure}{0.70\textwidth}
  \centering
  \includegraphics[width=1.0\linewidth]{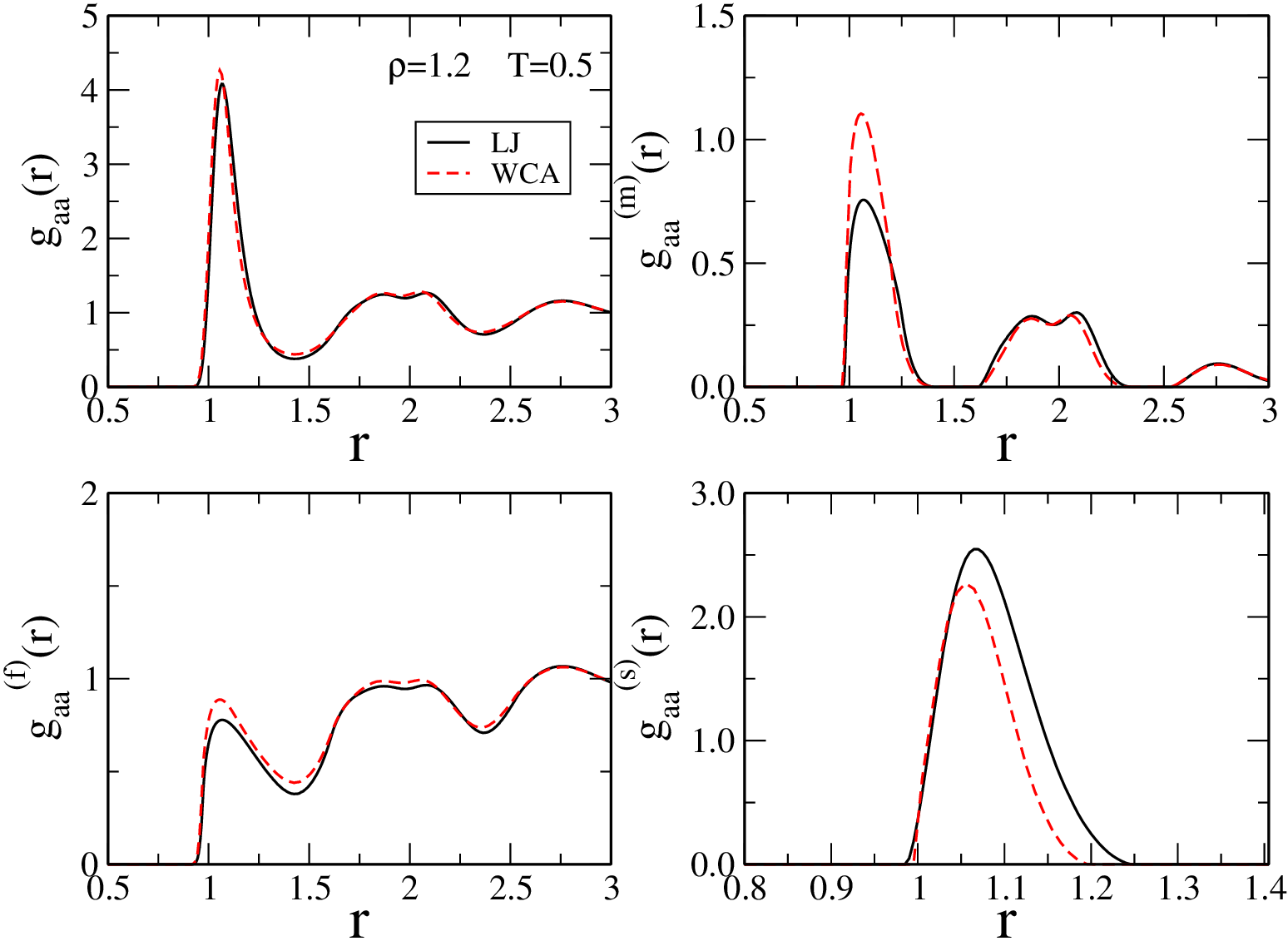} 
    \caption{}
\end{subfigure} \\%
\vspace{1cm}
\begin{subfigure}{0.70\textwidth}
  \centering
  \includegraphics[width=1.0\linewidth]{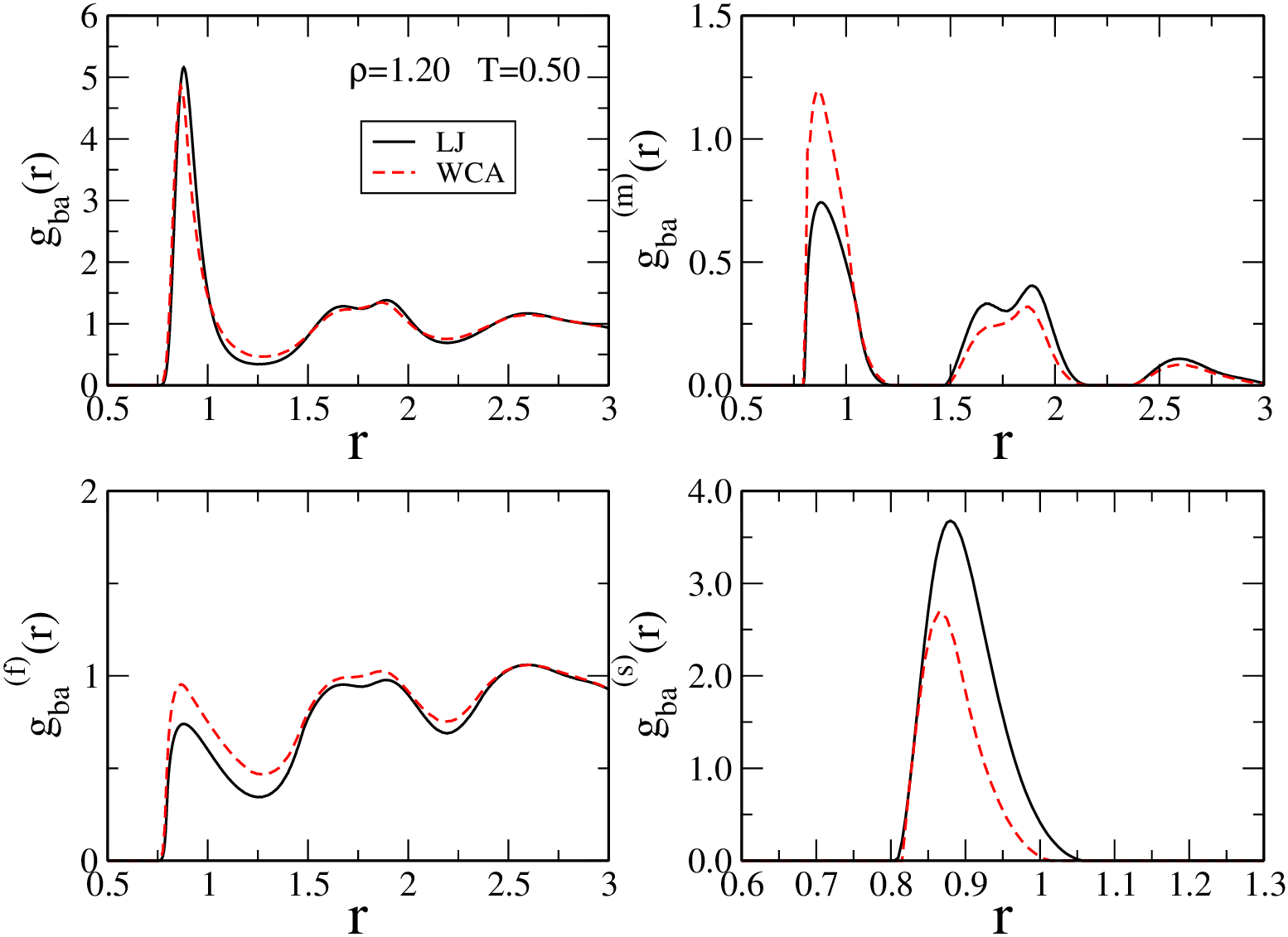}
  \caption{}
\end{subfigure}
\caption{Comparison of values of $g_{\alpha\gamma(r)}$ and its different parts $g_{\alpha \gamma}^{(f)}(r)$, $g_{\alpha \gamma}^{(m)}(r)$ and $g_{\alpha \gamma}^{(s)}(r)$ of the LJ
(full lines) and WCA (dashed lines) for $\rho=1.2$ and $T=0.5$.
In figure $(a)$, $\alpha,\gamma=a,a$ and in $(b)$, 
$\alpha,\gamma=a,b$. A large difference can be seen between values of
$g_{\alpha \gamma}^{(s)}(r)$ and of $g_{\alpha \gamma}^{(m)}(r)$ of the two systems, while values of $g_{\alpha\gamma}(r)$ are nearly identical.}
\label{fig-3}
\end{figure*}

\begin{figure*}
\begin{subfigure}{0.7\textwidth}
  \centering
  \includegraphics[width=1.0\linewidth]{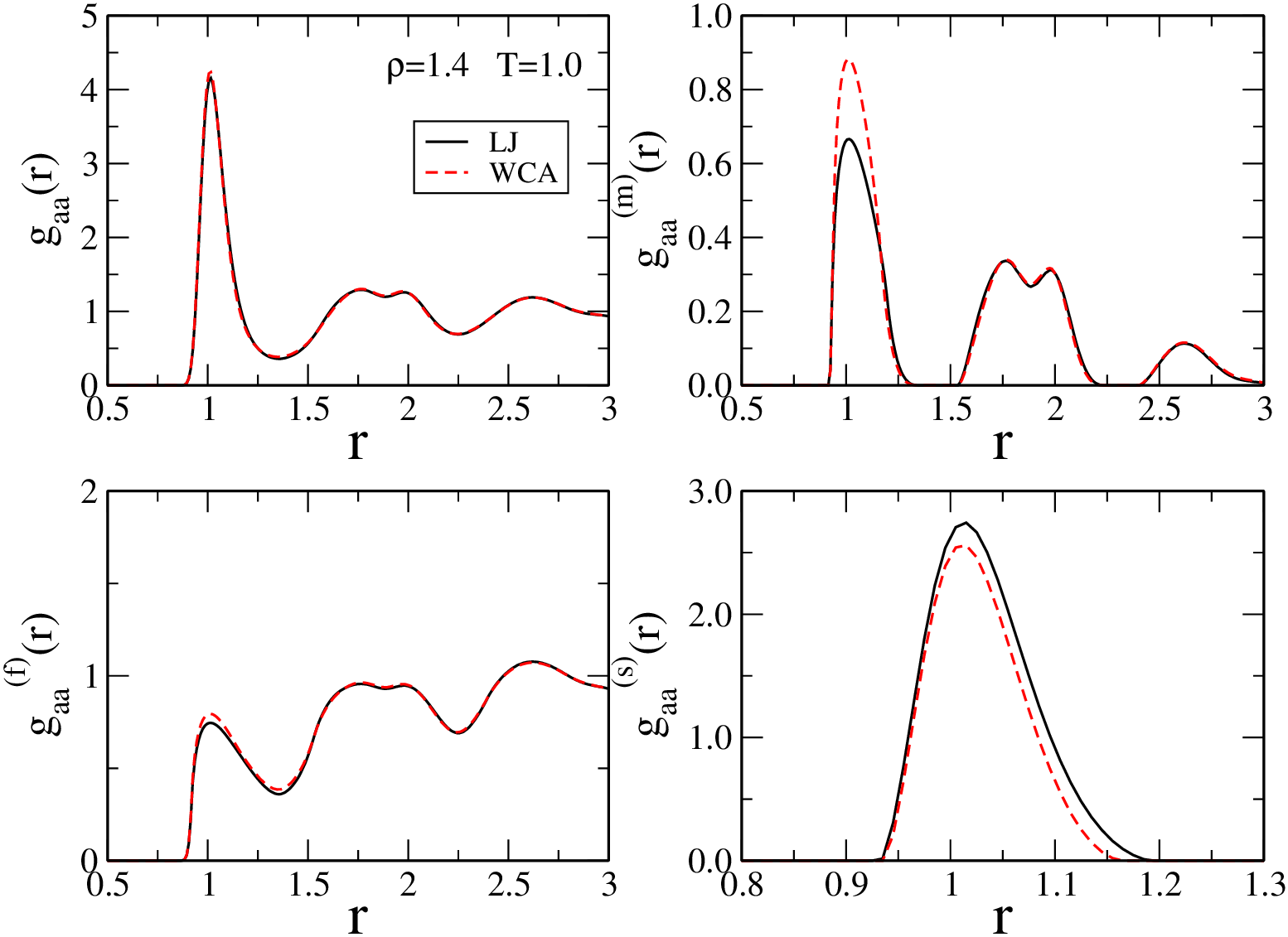}
  \caption{}
\end{subfigure} \\%
\vspace{1cm}
\begin{subfigure}{0.7\textwidth}
  \centering
  \includegraphics[width=1.0\linewidth]{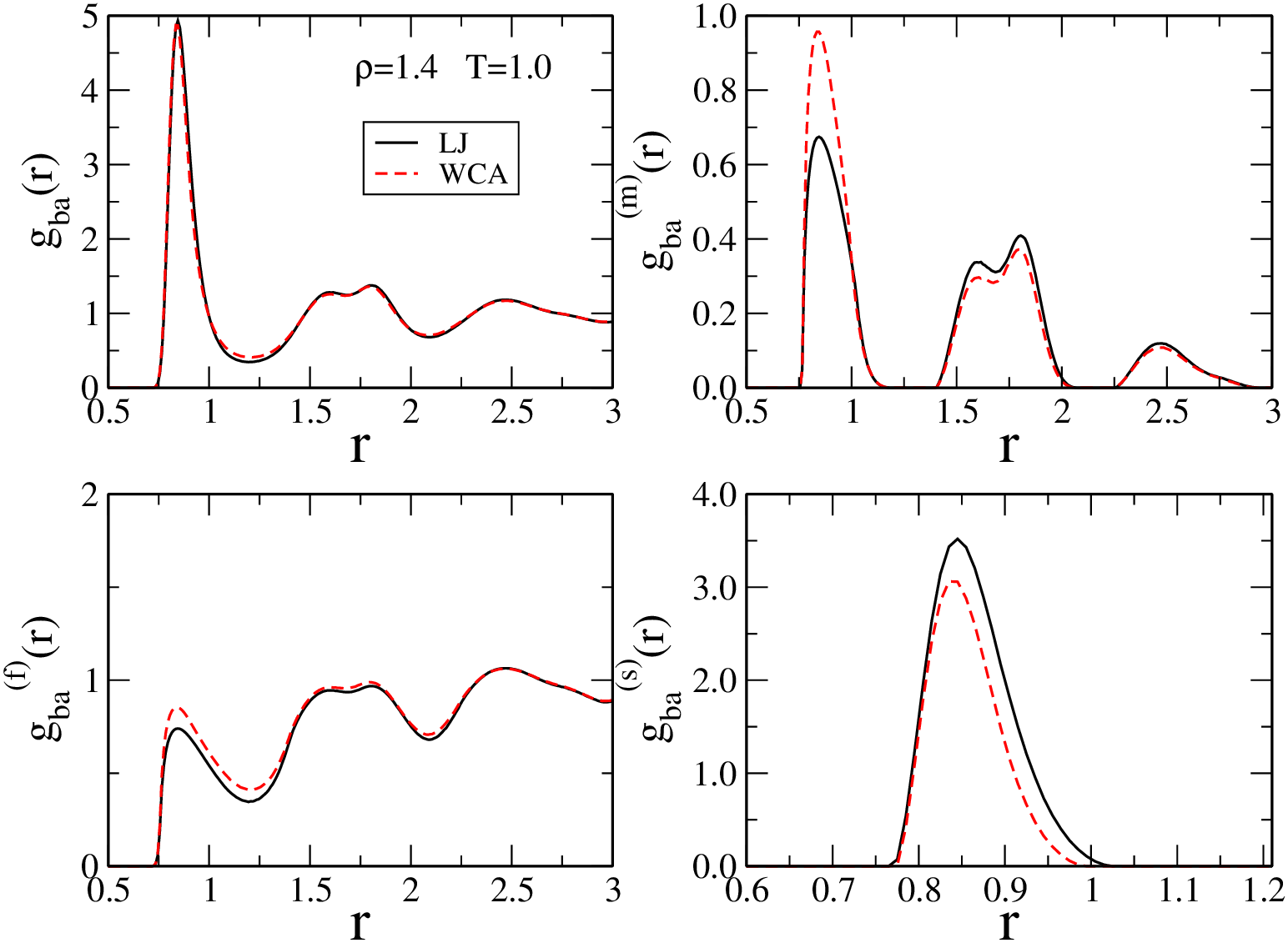}
  \caption{}
\end{subfigure}
\caption{Same as in Fig.~\ref{fig-3} but for $\rho=1.4$ and $T=1.0$. A relatively small difference is seen between values of $g_{\alpha \gamma}^{(s)}(r)$ and of $g_{\alpha \gamma}^{(m)}(r)$ of the two
systems compared to $\rho=1.2$.}
\label{fig-4}
\end{figure*}

\begin{figure*}
\begin{subfigure}{0.7\textwidth}
  \centering
  \includegraphics[width=1.0\linewidth]{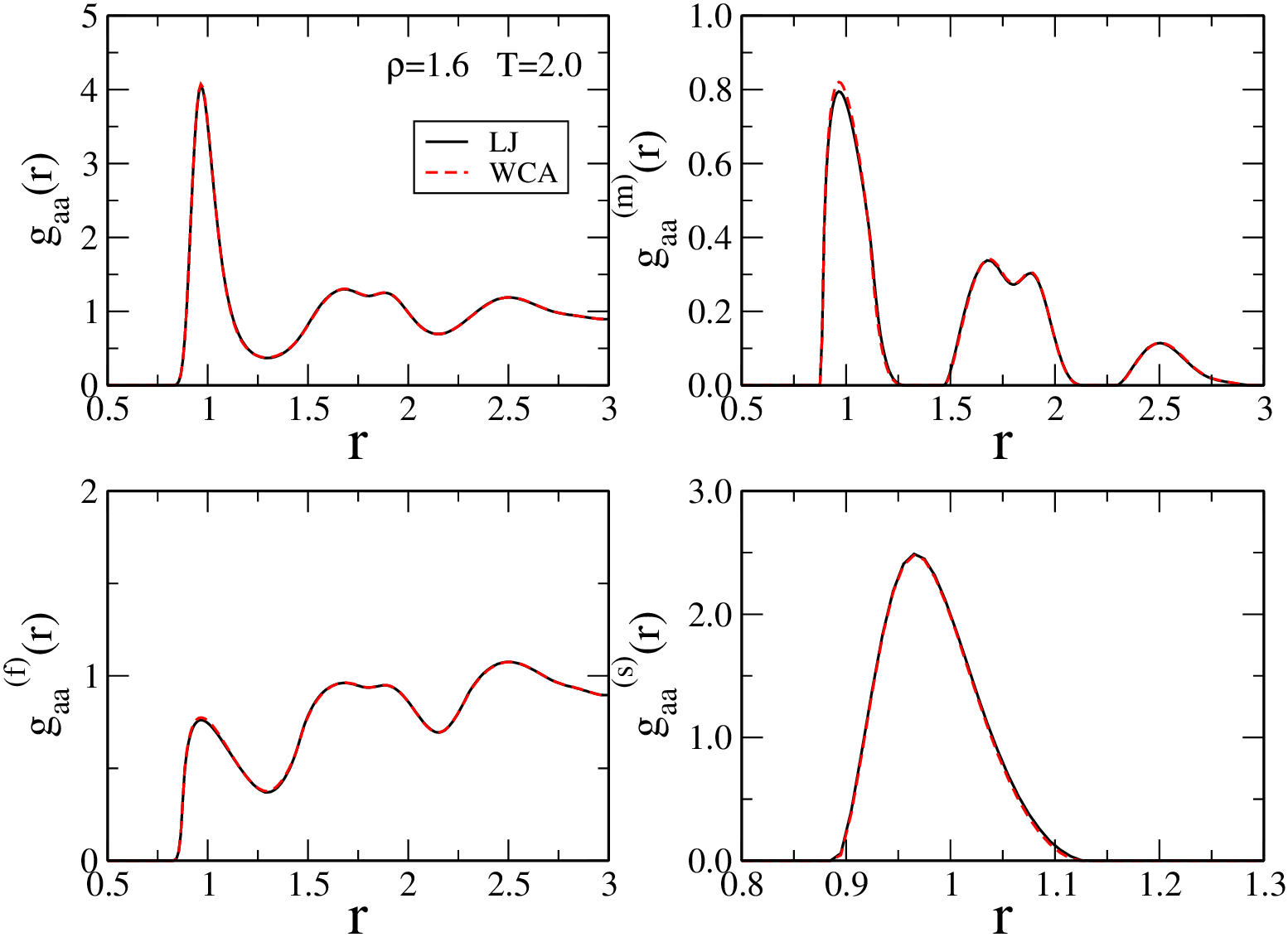}
  \caption{}
\end{subfigure} \\%
\vspace{1cm}
\begin{subfigure}{0.7\textwidth}
  \centering
  \includegraphics[width=1.0\linewidth]{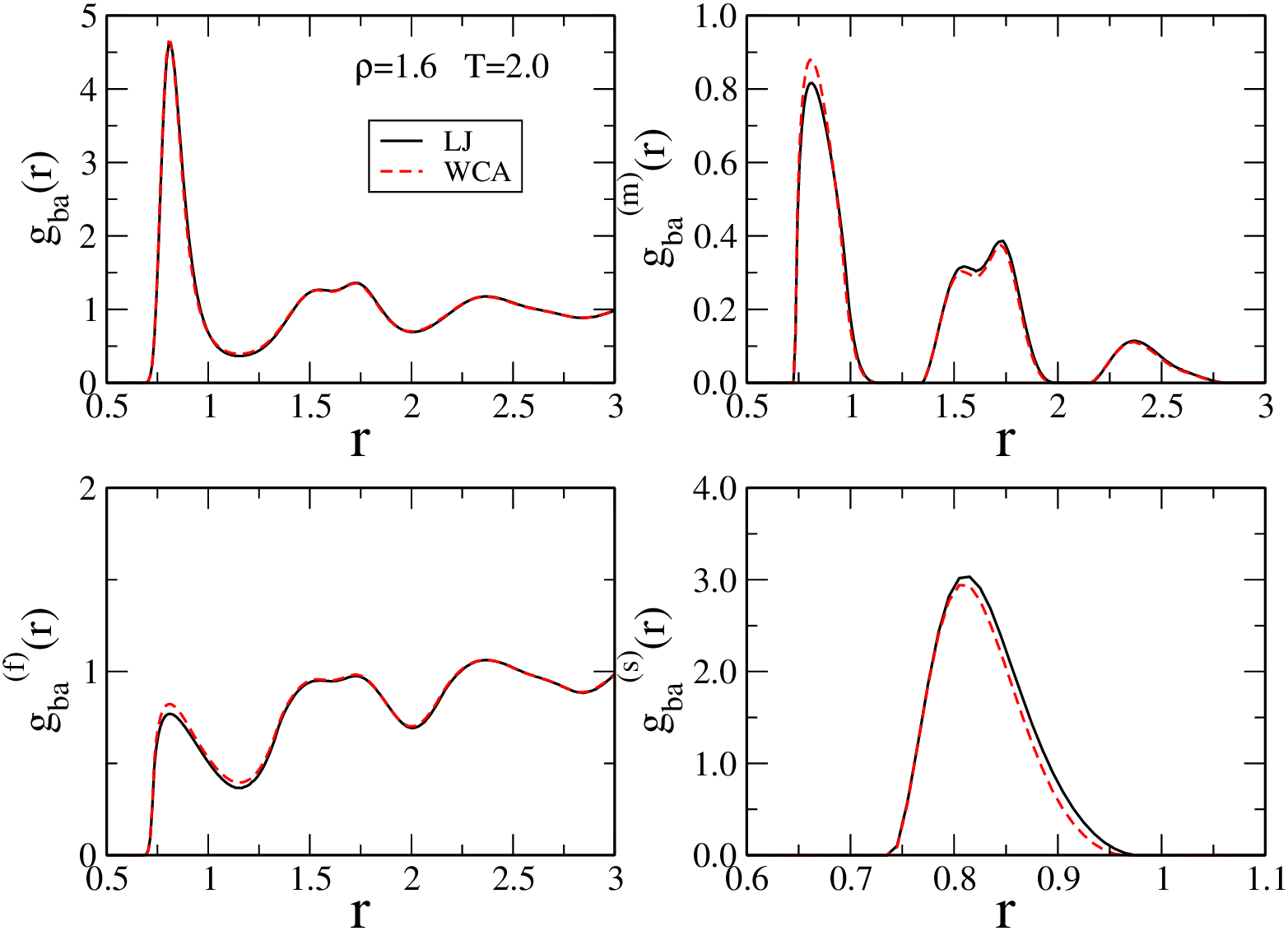}
  \caption{}
\end{subfigure}
\caption{Same as in Fig.~\ref{fig-3} but for $\rho=1.6$ and $T=2.0$. Values of $g_{\alpha\gamma}(r)$ and of its components, $g_{\alpha \gamma}^{(f)}(r)$, $g_{\alpha \gamma}^{(m)}(r)$ and $g_{\alpha \gamma}^{(s)}(r)$ of the two systems are nearly identical.}
\label{fig-5}
\end{figure*}

\begin{figure*}[t]
\includegraphics[scale=0.6]{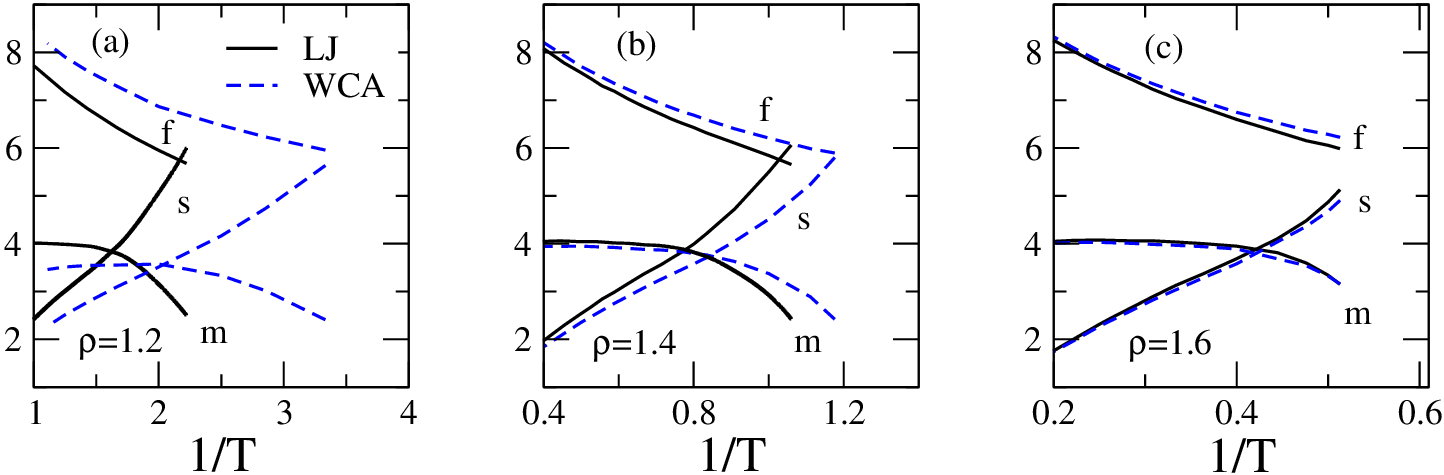}
\caption{Comparison of values of $n_{1}^{(f)}$, $n_{1}^{(m)}$ and  $n_{1}^{(s)}$ as a function of $1/T$ for the two 
systems. Letters $f$, $m$ and $s$ stand, respectively, for free, metastable and stable and the subscript $1$
stands for the first shell. Full lines represent values of the LJ system and dashed lines of the WCA system.}
\label{fig-6}
\end{figure*}

\begin{figure*}[t]
\vspace{1.3cm}
\includegraphics[scale=0.6]{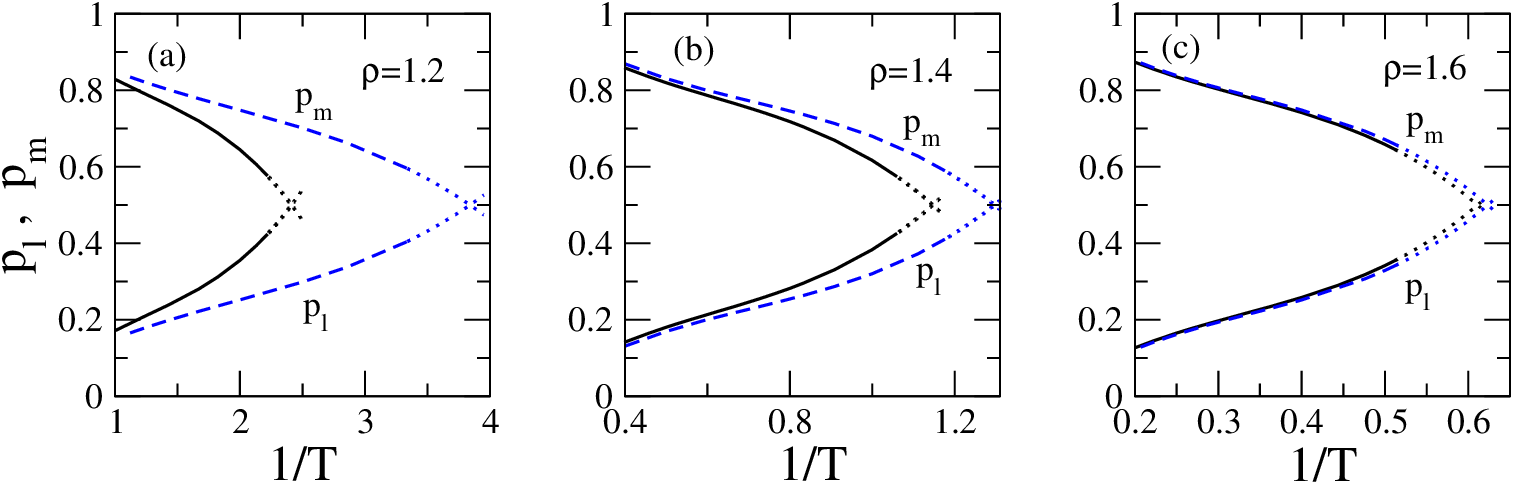}
\caption{Comparison of values of fraction of mobile, $p_{m}$, and immobile, $p_{l}$, (defined in the text) particles in the first shell around the central particle of the two systems. Line symbols are same as in Fig.~\ref{fig-6}. The dotted part of each 
line represents extrapolated values. The two curves of $p_{m}$ and $p_{l}$ meet at $T=T_{mc}$ where half of particles of the 
first neighbor become immobile.}
\label{fig-7}
\end{figure*}

\begin{figure}[t]
\includegraphics[scale=0.41]{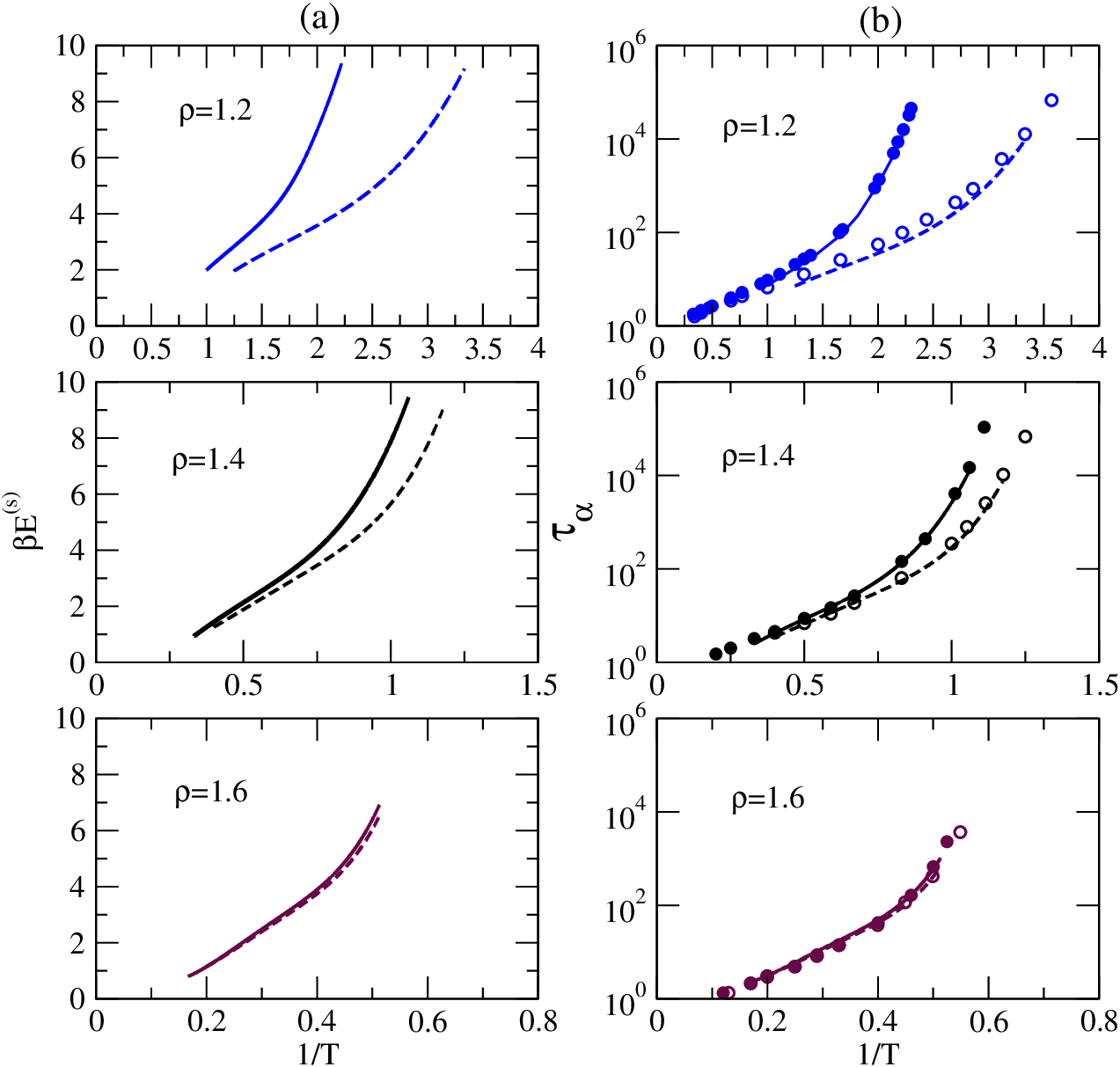}
\caption{Comparison of values of activation energy  $\beta E^{(s)}$ for the relaxation in column $(a)$
and the relaxation time in column $(b)$ as a function of $1/T$ at three densities of the two systems.
Curves (full for the LJ and dashed for the WCA) represent calculated values and circles 
(full for the LJ and open for the WCA) in column $(b)$ represent simulation values.}
\label{fig-8}
\end{figure}

\begin{figure}[]
\vspace{0.5cm}
\includegraphics[scale=0.27]{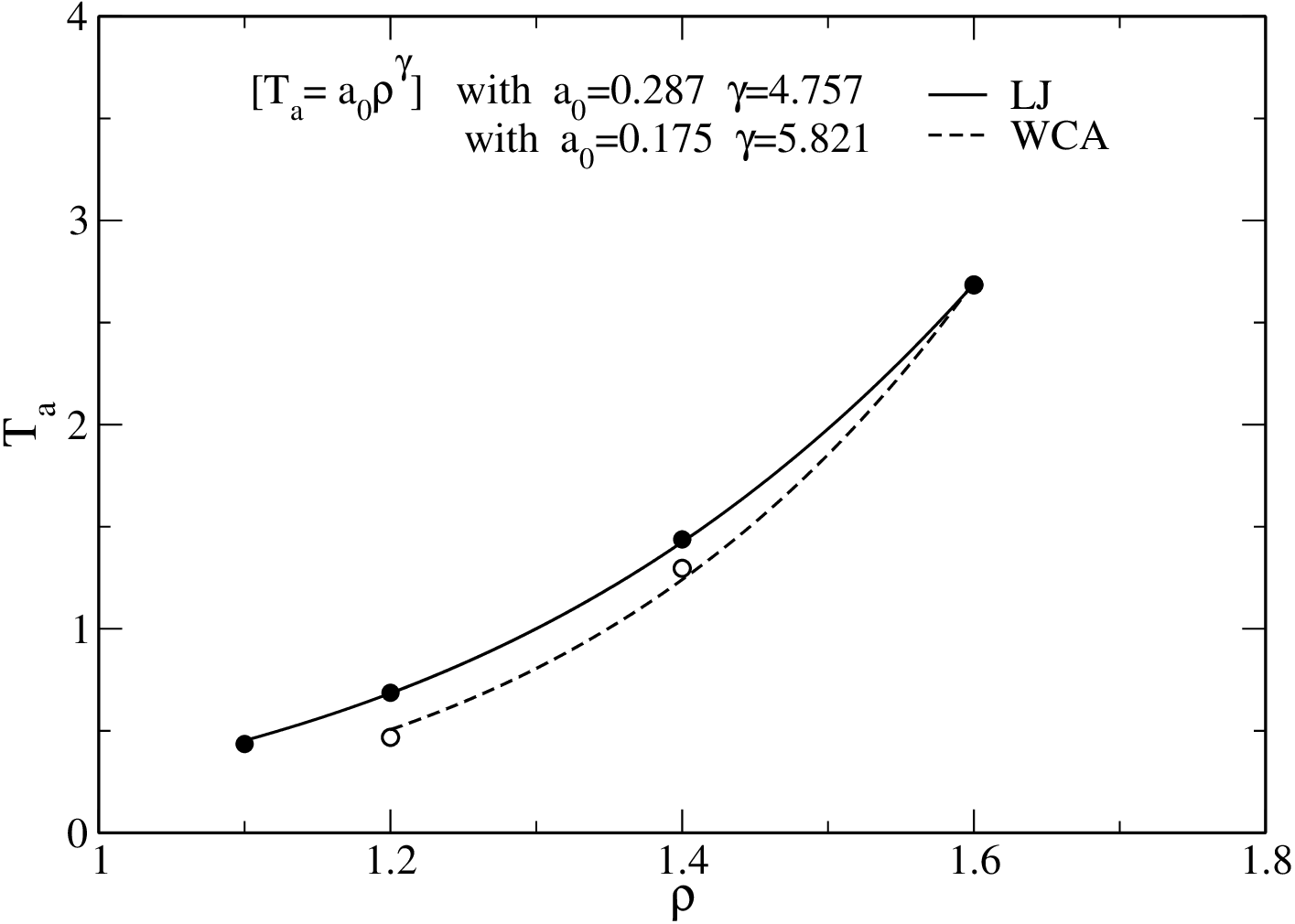}
\caption{Comparison of density dependence of the temperature $T_{a}$ of the two systems. Circles
(full for the LJ and open for the WCA systems) represent calculated values (given in Table~\ref{LJ-Data}) 
and curves (full for the LJ and dashed for the WCA) represent fit $T_{a}=a_{0}\rho^{\gamma}$ with 
$\gamma=4.757$ and $5.821$, respectively for the LJ and WCA systems.}
\label{fig-9}
\end{figure}

\begin{figure}[]
\includegraphics[scale=0.6]{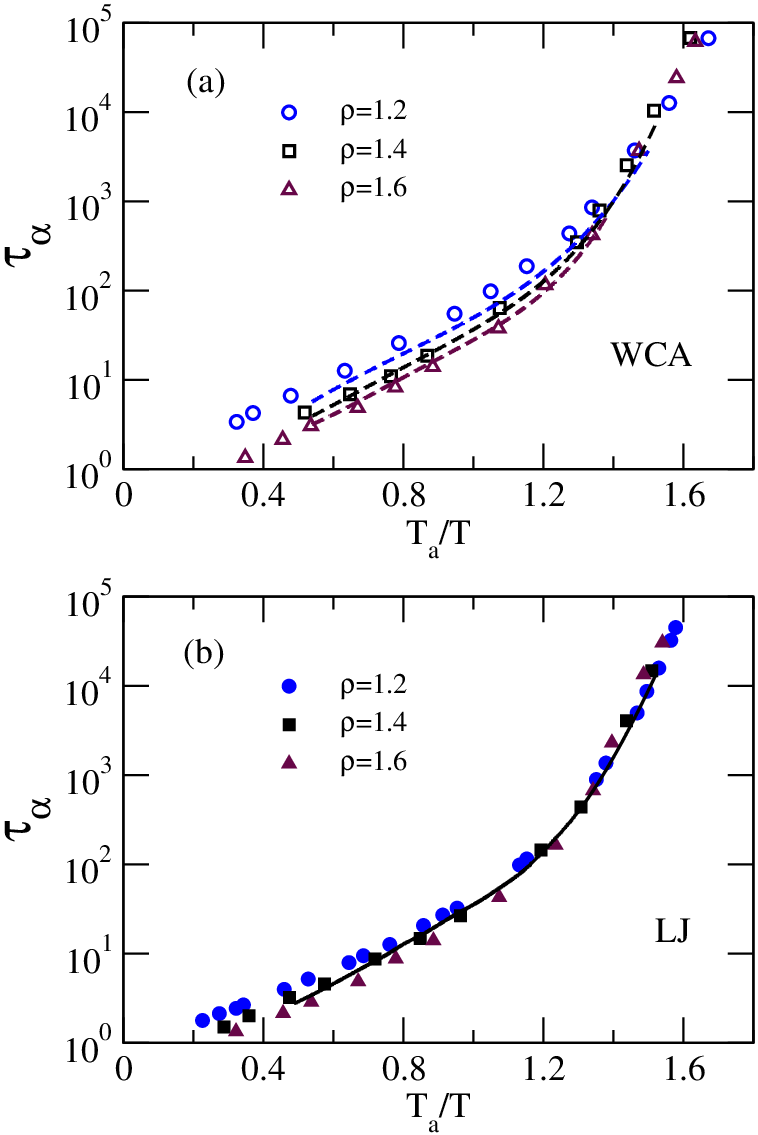}
\caption{Comparison of collapse of calculated (lines) and simulation (symbols) data of $\tau_{\alpha}$ of
the two systems as a function of $T_{a}/T$. While in the case of LJ system $(b)$ good collapse takes place,
in case of the WCA $(a)$, both the calculated as well as simulation values fail to collapse on one curve.}
\label{fig-10}
\end{figure}

In this section we report results found for the WCA system and compare them with those reported in $I$
for the LJ system at densities $\rho=1.2$, $1.4$ and $1.6$.

In Fig.~\ref{fig-1} values of $K$ calculated from Eqs.~$(4)-(7)$ with value of parameter $\psi(T)$ fixed at one
at all temperatures $T$, are plotted as a function of $T$. In the figure
we also plot values of $K$ found in a similar way (see Fig.~$2$ of $I$) for 
the LJ system.
We note that, in general, the temperature dependence 
of $K$ of the WCA system is similar to the one found for the LJ system; $K$ is constant above a temperature denoted as $T_{a}$
and decreases from its constant value on cooling below $T_{a}$. As explained in $I$, the decrease in value
of $K$ below $T_{a}$ is due to the unphysical condition that has been imposed on $\psi(T)$ by fixing its 
value equal to one for $T<T_{a}$. It is the temperature
independent value of $K$, \textit{i.e.} value of $K$ found for $T>T_{a}$, that should be, as argued in $I$,
used in Eq.~(\ref{Sc}). Values of $K$ and $T_{a}$ of both systems for three densities are listed in Table~\ref{LJ-Data}.
We note that value of $K$ for the WCA system is higher at the same density than that for the LJ system.
The difference, however, is found to decrease on increasing the density; while at $\rho=1.2$ the difference 
is $1.70$, it reduces to 0.25 at $\rho=1.6$. This density dependent difference is
a measure of the density dependent role of attractive interaction on configurations that is assessed by a CRC.

Values of $K$ listed in Table~\ref{LJ-Data} and of the configurational entropy $S_{c}$ found from simulations \cite{ABanerjee}
are used in Eq. (\ref{Sc}) to calculate values of $n^{(s)}(T,\rho)$. From known values of  $n^{(s)}(T,\rho)$
we determine $\psi(T,\rho)$ from Eqs.~$(4)-(6)$ for different values of $T$ and $\rho$.
We plot $\psi(T)$ vs $T$ in Fig.~\ref{fig-2}. We note that for both systems, the high temperature value of $\psi(T)$
is equal to one and decreases rather sharply on cooling below $T_{a}$. The transition from the high temperature 
region to the low temperature region begins at $T_{a}$ with a crossover region spreading over a narrow
temperature width. Since the crossover region 
separates the high temperature region where slowing down of dynamics is slower from the low temperature region 
where slowing down of dynamics is faster, $T_{a}$ can be taken as the crossover temperature. From Fig.~\ref{fig-2} 
one finds that for $\rho=1.2$
dynamics of the LJ system would start slowing down at faster rate for $T \leq 0.68$ whereas the slowing down 
of the WCA system would remain at slower rate till $T_{a}=0.47$ on cooling. This results into a large difference in values 
of $\tau_{\alpha}$ below $T=0.68$ of the two systems. However, as $\rho$ increases temperature dependence of dynamics of the two systems
come closer as the difference in values of $T_{a}$ decreases.

In Figs.~$(3-5)$ we compare radial distribution function $g_{\alpha \gamma}(r)$ and its different parts, 
$g_{\alpha \gamma}^{(f)}(r)=g_{\alpha \gamma}(r)-g_{\alpha \gamma}^{(b)}(r)$, 
$g_{\alpha \gamma}^{(m)}(r)=g_{\alpha \gamma}^{(b)}(r)-g_{\alpha \gamma}^{(s)}(r)$ and $g_{\alpha \gamma}^{(s)}(r)$,
where $\alpha, \gamma=a, a$ and $a, b$ (notations are of $I$) for the two systems at $(\rho,T)=(1.2, 0.5), (1.4, 1.0)$ and
$(1.6, 2.0)$, respectively. As defined in Sec.~\ref{Theory}, $g_{\alpha \gamma}^{(f)}(r)$, $g_{\alpha \gamma}^{(m)}(r)$
and $g_{\alpha \gamma}^{(s)}(r)$ describe, respectively, distributions of free, $m$ and $s$ -particles around a central
particle in a liquid. A look at these figures shows that while $g_{\alpha \gamma}(r)$ of the two systems in all cases 
nearly overlap, differences are seen in values of $g_{\alpha \gamma}^{(f)}(r)$, $g_{\alpha \gamma}^{(m)}(r)$
and $g_{\alpha \gamma}^{(s)}(r)$. In particular, first peaks of $g_{\alpha \gamma}^{(m)}(r)$
and $g_{\alpha \gamma}^{(s)}(r)$ in Fig.~\ref{fig-3} show large difference at $\rho=1.2$. The difference becomes less 
pronounced at $\rho=1.4$ (Fig.~\ref{fig-4}) and almost disappears at $\rho=1.6$ (Fig.~\ref{fig-5}). In all cases, the peak of 
$g_{\alpha \gamma}^{(m)}(r)$ is higher and $g_{\alpha \gamma}^{(s)}(r)$ is lower for the WCA system compared to
those for the LJ system. Since free and $m$ particles are mobile whereas $s$ particles are immobile, a particle 
in the WCA system is surrounded by a relatively larger population of mobile particles and a less population of immobile
neighbors compared to those of the LJ system, leading to difference in their dynamical behavior.
The subtle local structural order that defines the cooperativity of relaxation and remains hidden to 
experiments that measure pair correlation functions is defined in terms of $g_{\alpha \gamma}^{(s)}(r)$.

A useful information that shed light on the underlying features related to the local structural order and
dynamics can be derived from the dynamical states of particles of the first shell surrounding
the central particle as a function of temperature and density. In Fig.~\ref{fig-6} we compare values of 
$n_{1}^{(f)}$, $n_{1}^{(m)}$ and $n_{1}^{(s)}$ as a function of $1/T$ for the two systems at different densities. Here the subscript $1$
indicates the first coordination shell. In the figure solid lines indicate values for the LJ system and dashed lines
for the WCA system. Letters $f$, $m$ and $s$ stand, respectively, for free, metastable and stable. 
We note that temperature dependence of number of particles of different dynamical states that 
surround the central particle in the two systems is, in general, similar. In both cases we find 
that at high temperatures most particles are free while a few are $m$-particles and a very few are
$s$-particles. On cooling the systems, $n_{1}^{(m)}(T)$ remains nearly constant but $n_{1}^{(s)}(T)$
increases though slowly till $T=T_{a}$, but for $T<T_{a}$, $n_{1}^{(m)}(T)$ decreases and $n_{1}^{(s)}(T)$
increases with increasing rate at the cost of both free and $m$-particles.
The rate is expected to increase rapidly on further lowering of temperatures, resulting into a rapid
increase in number of $s$-particles and therefore in the cooperativity of relaxation. There is,
however, a large difference particularly at low temperatures, in the number of particles of a given
category of the two systems at $\rho=1.2$; the difference decreases on increasing the
density and almost disappears at $\rho=1.6$.

One may prefer to use fraction of mobile (free plus $m$-particles) and immobile (localized) first neighbors defined, respectively, as 
$p_{m}(T)=\frac{n_{1}^{(f)}(T)+n_{1}^{(m)}(T)}{n_{1}^{(t)}(T)}$ and 
$p_{l}(T)=\frac{n_{1}^{(s)}(T)}{n_{1}^{(t)}(T)}$ to compare the local ordering relevant
to dynamics. In Fig.~\ref{fig-7} we plot and compare values of 
$p_{m}(T)$ and $p_{l}(T)$ of the two systems as a function of $1/T$.
In the figure, lines (full for the LJ and dashed for the WCA) represent
calculated values and dotted part of each line represents extrapolated
values. A glance at this figure reveals how the local ordering defined in
terms of the fraction of mobile and immobile particles around the central
particle in the two systems compare with each other at different densities.
The extrapolated parts of lines representing $p_{m}(T)$ and $p_{l}(T)$
meet at a temperature (denoted as $T_{mc}$) where half of the neighbors
become immobile. The values of $T_{mc}$ found for densities $\rho=1.2$,
$1.4$ and $1.6$ for the LJ system are, respectively, $0.41$, $0.87$
and $1.64$ whereas the corresponding values for the WCA system are
$0.26$, $0.77$ and $1.62$. The large difference in the temperature
dependence of $p_{m}(T)$ and $p_{l}(T)$ and values of $T_{mc}$ found
at $\rho=1.2$ that decreases on increasing the density further explains why
dynamics of the two systems show large difference at $\rho=1.2$ but become almost
identical at $\rho=1.6$.

The effective activation energy $\beta E^{(s)}(T,\rho)$ and the relaxation time $\tau_{\alpha}(T, \rho)$
calculated, respectively, from Eqs.~(\ref{energy}) and  (\ref{tau_E}) are plotted in Fig.~\ref{fig-8}. In Fig.~\ref{fig-8}$(a)$
values of $\beta E^{(s)}$ of the two systems are compared as a function of $1/T$. In Fig.~\ref{fig-8}$(b)$
we compare values of $\tau_{\alpha}/\tau_{0}$ of the two systems with each other and with values found from
simulations \cite{Tarjus_PRL,Tarjus_JCP}. An excellent agreement is found with simulation values for all densities for both systems.

The density dependence of $T_{a}$ is shown in Fig.~\ref{fig-9}. In the figure, circles - full for the LJ and open 
for the WCA- denote calculated values and the curves- full line for the LJ and dashed line for the 
WCA- represent fit with a power law form $T_{a}=a_{0}\rho^{\gamma}$. In case of the LJ system,
$a_{0}=0.287$ and $\gamma=4.757$ while in case of the WCA, $a_{0}=0.175$ and $\gamma=5.821$.
It was shown in $I$ that when $\psi(T)$, $\tau_{\alpha}$, $\beta E^{(s)}$, etc., are plotted as a 
function of $T_{a}/T$ (or $T/T_{a}$)the data collapse on master curves. However, in case of the WCA such a collapse 
fails in agreement with result found from simulations \cite{Tarjus_PRL,Tarjus_JCP}. We plot calculated values (shown by lines) and simulation
values by symbols of $\tau_{\alpha}/\tau_{0}$ as a function of $T_{a}/T$ for densities $\rho=1.2$,
$1.4$ and $1.6$ in Fig.~\ref{fig-10}$(a)$ for the WCA system and in \ref{fig-10}$(b)$ for the LJ system. 
While in case of the LJ system a good collapse of data happens, in case of the WCA both the 
calculated as well as simulation values fail to collapse on one curve and therefore violates
the density scaling relation; a feature attributed to the fact that the WCA liquid follows an ``isomorph''
different from that of the LJ liquid \cite{Toxvaerd,Pedersen}.


\section{Conclusion \label{Conclusion}}
Through an extensive comparison of the behavior of a LJ glass forming liquid and its WCA reduction to a model 
with truncated potential without attractive tail, we have shown that our theory brings out several underlying 
features that characterize slowing down of dynamics of these systems. 
It is shown that though the equilibrium static structures measured by the pair correlation functions, $g_{\alpha \gamma}(r)$
of the two systems are nearly identical, there is a marked difference, particularly at low densities and low 
temperatures, in their components representing free, metastable and stable particles distributed in 
coordination shells around a central particle.
In particular, the parameters $p_{m}(T)$ and $p_{l}(T)$ which define, respectively, the fraction of mobile
and immobile first neighbors and plotted in Fig.~\ref{fig-7} provide information about the local 
structure relevant to dynamics.

The other intrinsic parameters stemming from the intermolecular interactions and which explain why 
slowing down of dynamics of the two systems are qualitatively and quantitatively different at lower densities
and lower temperatures but nearly identical at higher densities and higher temperatures are $\psi(T)$
and the crossover temperature $T_{a}$. The value of parameter $\psi(T)$ which measures the effect of fluctuations
embedded in the system on stabilizing the shape and size of CRC, takes a turn from its high temperatures
value of $1$ at $T=T_{a}$ and starts decreasing rather sharply as $T$ is lowered. There is a one-to-one
correspondence between the crossover region of $\psi(T)$ and $\tau_{\alpha}$ (for more details see $I$)
indicating the significance of embedded fluctuations on temperature and density dependence of local
structure and dynamics. The temperature $T_{a}$ separates the high temperature behavior region where 
slowing down of dynamics is slower from a low temperatures region where slowing down of dynamics is faster.
Both $\psi$ and $T_{a}$ depend on details of intermolecular interactions.
Whenever the attractive part of intermolecular interaction is effective in suppressing entropy driven escapes of particles out of shells, the crossover temperature $T_{a}$ shifts to higher temperatures resulting 
into slowing down of dynamics at faster rate than in absence of the attractive part.
The reason why value of $\tau_{\alpha}$ for $\rho=1.2$ is much smaller in the WCA system,
say at $T=0.5$, than the LJ ones, lies in the difference in their values of $T_{a}$ (see Table~\ref{LJ-Data}).
The quantitative and qualitative agreements found with simulation results
for the two systems suggest that our theory accurately describes the intricate nature of the connection between the local structural order and dynamics arising due to attractive and repulsive interactions in glass-forming liquids.

\section*{Acknowledgments}
We thank Dr. S. M. Bhattacharyya for helpful discussions and for providing simulation data of pair correlation functions $g_{\alpha\gamma}(r)$ and configurational entropy $S_{c}$.
One of us (A.S.) acknowledges research fellowship from the Council of Scientific and Industrial Research, New Delhi, India.


\end{document}